# Towards Understanding of Frequency Dependence on Sound Event Detection

Hyeonuk Nam, Seong-Hu Kim, Deokki Min, Byeong-Yun Ko and Yong-Hwa Park

***Abstract*— In this work, we conduct an in-depth analysis of two frequency-dependent methods for sound event detection (SED): FilterAugment and frequency dynamic convolution (FDY conv). The goal is to better understand their characteristics and behaviors in the context of SED. While SED has been rapidly advancing through the adoption of various deep learning techniques from other pattern recognition fields, such adopted techniques are often not suitable for SED. To address this issue, two frequency-dependent SED methods were previously proposed: FilterAugment, a data augmentation randomly weighting frequency bands, and FDY conv, an architecture applying frequency adaptive convolution kernels. These methods have demonstrated superior performance in SED, and we aim to further analyze their detailed effectiveness and characteristics in SED. We compare class-wise performance to find out specific pros and cons of FilterAugment and FDY conv. We apply Gradient-weighted Class Activation Mapping (Grad-CAM), which highlights time-frequency region that is more inferred by the model, on SED models with and without frequency masking and two types of FilterAugment to observe their detailed characteristics. We propose simpler frequency dependent convolution methods and compare them with FDY conv to further understand which components of FDY conv affects SED performance. Lastly, we apply PCA to show how FDY conv adapts dynamic kernel across frequency dimensions on different sound event classes. The results and discussions demonstrate that frequency dependency plays a significant role in sound event detection and further confirms the effectiveness of frequency dependent methods on SED.**

## I. Introduction

Sound event detection (SED) aims to detect the time localization (onset and offset) of desired sound event classes within given an audio clip [1]–[4]. It finds its application in various areas such as robotics [2], automation [2], public/private monitoring [1], [3], machine diagnosis [2], multimedia information retrieval [1], [3], etc. Recently, SED has benefited from the rapid development of deep learning (DL) methods from major pattern recognition fields such as computer vision (CV) [5]–[10], natural language processing (NLP) [1], [11], [12], and speech recognition [5], [13], [14]. These fields were benefitted from powerful DL based pattern recognition with data-driven techniques [15]–[21]. DL methods from CV applied to 2D image data have been widely adopted in audio and speech DL applications to process 2D audio data [22]–[24]. In addition, DL methods from NLP processing word sequences have been adopted to process the temporal sequenced audio data [15]–[18], [25]. However, DL methods from other fields are often applied to SED without proper adaptation, potentially introducing biases that degrade performance [12], [14], [26], [27]. To ensure the robustness of SED, we need a thorough examination on the difference between SED and the original field of the DL methods, leveraging insights from the acoustics and signal processing domain knowledge [26].

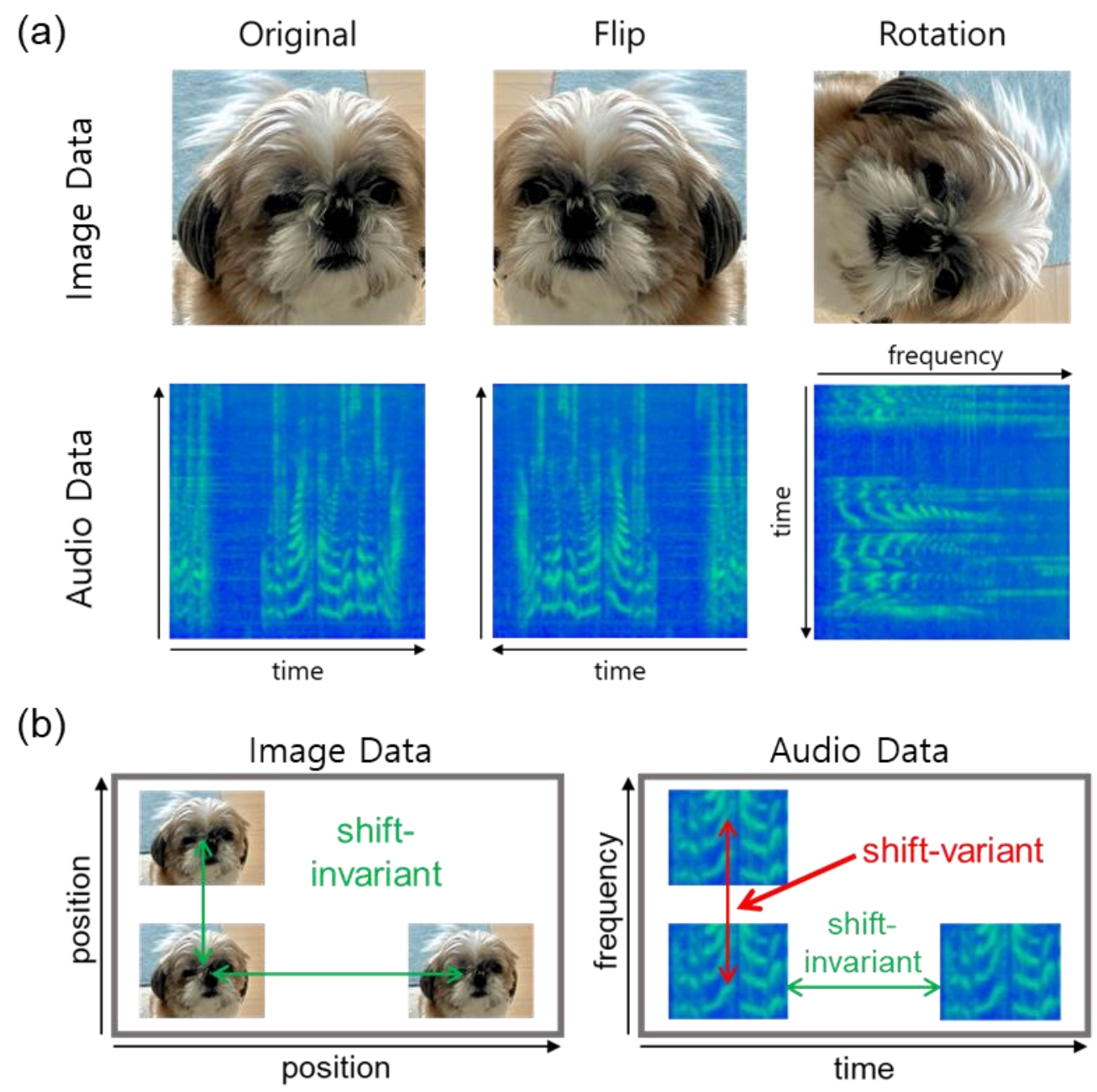

**Fig. 1.** An illustration of the physical difference between image data domain and 2D domain data. The difference is illustrated in relation to (a) data augmentation and (b) 2D convolution module.

Recent advances in SED have been driven by deep learning, with numerous transformer-based architectures demonstrating strong performance. These include dual-encoder models that combine CNN and Transformer modules [28]–[32], transformer-RNN hybrid [33], and fully transformer-based network [34]. However, fully transformer-based models have not consistently outperformed dual-encoder models [30], [31], [34]. This suggests that the ability of CNNs to capture local spectral patterns remains crucial for effective SED, complementing the transformer's capacity to model global dependencies. As a result, CNNs are still actively used in state-of-the-art SED systems, often as front-end encoders.

Frequency is one of the important aspects of 2D audio data that requires thorough investigation. Frequency components exhibit characteristic patterns of sound events over different frequency ranges. Frequency dimension visually presents us sound patterns by displaying change in frequency components over time [22]. Visualization of sound patterns allowed us to adopt DL methods from computer vision field to the audio and speech domains. However, there has not been sufficient investigation on the difference between time and frequency dimensions when applying DL methods from computer vision. Unlike image data with two dimensions of the same physical quantity (position), 2D audio data has different physical quantities (time and frequency). Due to the difference between time and frequency dimensions and causality of time dimension, data augmentation methods for image data such as flip and rotation ruins information in 2D audio data, as illustrated in Fig. 1. (a). Also, unlike position and time, frequency is shift-variant dimension, as illustrated in Fig. 1. (b). In other words, shifting a time-frequency pattern along the frequency axis alters the perceived sound, as frequency represents a non-uniform, semantically meaningful dimension. To address these challenges, we previously proposed two frequency-dependent methods on SED. First, FilterAugment is a data augmentation technique that amplifies or attenuate random frequency bands during training to train SED models to be robust to various acoustic environments [35]. Second, frequency dynamic convolution (FDY conv) applies convolution kernels that adapts to each frequency bins' contents to release translation equivariance of 2D convolution along the frequency axis [26]. These methods were proven to significantly improve SED performance.

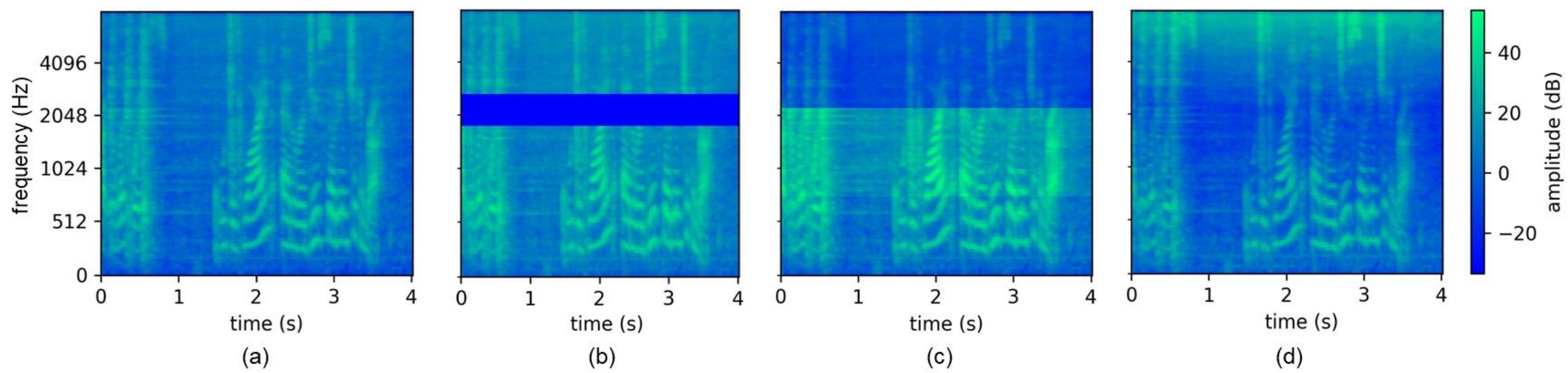

**Fig. 2.** Examples of applying frequency-wise data augmentations to a mel spectrogram of a speech sound example. (a) displays the original spectrogram, (b) demonstrates the result of applying frequency masking, (c) shows the result of applying step type FilterAugment, where three frequency bands are modified by attenuating the high band, amplifying the middle band, and slightly amplifying the low band, and (d) presents the result of applying linear type FilterAugment, with amplification on the highest frequency bin, attenuation on the middle frequency bin, and slight amplification on the lowest frequency bin with intermediate frequency bins linearly interpolated.

FilterAugment and FDY conv have demonstrated clear utility in SED tasks. FilterAugment has been adopted as a data augmentation strategy in many state-of-the-art SED models [28]–[32], [34], while CNN–Transformer dual encoder architectures have become common in high-performing systems [28], [30]–[32]. Some of these models incorporate FDY Conv in their CNN encoders [28], [32], and even in those that do not, the aforementioned issues related to applying 2D convolution on audio spectrograms still persist [29]–[31]. Therefore, FDY Conv—and its extensions such as DFD Conv and PFD Conv—remains a relevant and necessary component for addressing frequency-related challenges in CNN-based SED systems [32], [36], [37].

### *A. Motivation and Contributions*

While the aforementioned methods have demonstrated strong performance, there has been limited investigation into how they specifically impact SED performance and behavior. In this work, we further analyze the effectiveness and detailed characteristics of FilterAugment and FDY conv on SED to add insight on the specific behavior of the methods with pros and cons on SED. The main contributions of this paper follow:

1. Class-wise analysis on SED models with and without FilterAugment and FDY conv demonstrates effectiveness and weaknesses to various sound events.
2. Detailed discussion on different behaviors of frequency-wise data augmentations (two types of FilterAugment and frequency masking) was made utilizing Grad-CAM.
3. Simpler frequency-dependent convolution methods (FK conv and FW conv) were designed to comparatively analyze the strength of FDY conv and how frequency dependent kernels affect SED.
4. Principal component analysis (PCA) explains the behavior of attention weights in FDY conv by illustrating variation of frequency-adaptive kernel over frequency dimension.

## II. Frequency-Dependent Methods

The significance of the frequency dimension in audio and speech applications has been highlighted in recent works in the field of speaker recognition, achieving state-of-the-art performance [38]–[41]. A few recent studies have focused on addressing frequency dimension issues in SED [33], [42] with state-of-the-art performances as well. To examine the purpose and significance of methods that address frequency dimension, we discuss and analyze two methods that we previously proposed for SED in this study.

### *A. FilterAugment*

Data augmentation increases the size of the dataset by applying modification to data to provide various view [13], [35]. Audio and speech domains use audio signal processing methods such as adding noise, time stretching or pitch shifting for data augmentation purpose [43]–[45]. While these methods have been used in conventional audio applications such as music production, they are not necessarily optimal for training DL models and hard to use. To apply more straightforward data augmentation, SpecAugment directly applies time warping, time masking, and frequency masking on the log mel spectrogram [13]. Despite their simplicity, time and frequency masking showed great performance by removing data within specific time or frequency ranges on the mel spectrogram. Although such drastic changes can result in unnatural sounding data, they still effectively generalize DL models.

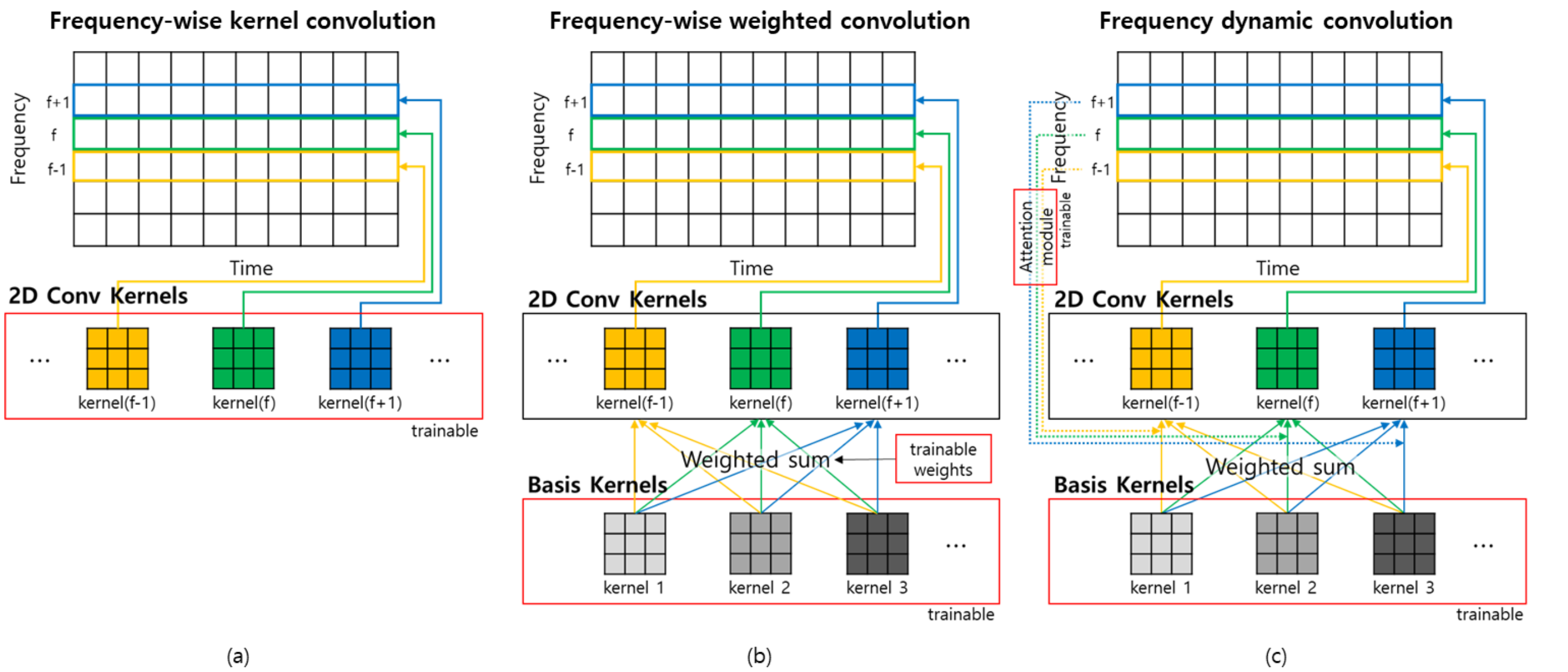

**Fig. 3.** An illustration of different approaches for 2D convolution with frequency-dependent kernels. (a) shows separately trained kernels on each frequency bin, (b) shows a kernel formed by a weighted sum of trained basis kernels using weights fixed on each frequency bin, and (c) shows a kernel formed by a weighted sum of trained basis kernels using weights inferred from the frequency bin. The parameters in the red boxes indicate that they are trainable parameters.

Similarly, FilterAugment was proposed in previous work [35]. *FilterAugment* is a data augmentation method designed to simulate various acoustic environments by applying random weights on frequency bands as illustrated in Fig. 2 [35]. It was motivated by the fact that the human auditory system robustly recognize sound events that occur in different acoustic environments, those distort frequency components of sound events by physical interactions [22], [35], [46]. To make SED models robust against distortion in the frequency components, FilterAugment applies random weights on the frequency bands those are formed by randomly setting frequency boundaries. Application of the weights differ by the two types of FilterAugment: step type and linear type. *Step type FilterAugment* applies step-like filters by multiplying constant weight on each frequency band, as illustrated Fig. 2(c) where horizontal lines on ~800 Hz and ~2100 Hz are the frequency boundaries showing the abrupt weight change across the frequency dimension. *Linear type FilterAugment* applies continuous filter by first assigning weights on each frequency boundary and then linearly interpolating weights between frequency boundaries as illustrated in Fig. 2(d), showing gradual and continuous change of weighting across frequency dimension.

FilterAugment is designed to extract sound event patterns from a wider frequency range by randomly amplifying and attenuating different frequency bands. However, previous work supports this presumption only by the performance improvement with FilterAugment. To further verify its effectiveness on SED, we applied Gradient-weighted Class Activation Mapping (Grad-CAM) [47] to SED models trained with and without FilterAugment and frequency masking, so called ‘frequency-wise data augmentations’ in this work.

### *B. Frequency dynamic convolution and frequency-wise convolutions*

2D convolution has been widely applied to audio and speech field as it effectively recognizes patterns within 2D audio data. However, 2D convolution applies inductive bias that is unsuitable for 2D audio data [26]. 2D convolution module is translation equivariant on both dimensions of image data. Translation equivariance refers to a property of a function, where translating the input results in a translated output.

$$T(f(x)) = f(T(x)) \tag{1}$$

where $T(x)$ refers to translation and $f(x)$ refers to any translation equivariant function. In other words, applying a translation after a function is equivalaent to applying the function after translation. This property is well-suited for image data, where both spatial dimensions (height and width) are shift-invariant—shifting the image content does not change its meaning. However, in 2D audio representations such as spectrograms, the two axes have different characteristics: The time axis is shift-invariant (event timing can vary), The frequency axis is not shift-invariant (shifting frequencies changes the meaning of the sound). Therefore, enforcing translation equivariance in both dimensions—especially in the frequency direction—is inappropriate for SED. Instead, it is more reasonable to relax translation equivariance along the frequency axis when applying 2D convolutions in SED models.

*Frequency dynamic convolution (FDY conv)* is proposed to address the issue of translational equivariance in 2D convolution along the frequency dimension by varying 2D convolution kernels to adapt to each frequency bin [26], [40]. Additionally, the attention mechanism which adapts convolution kernels to frequency bins further optimizes convolution kernels to given input. FDY conv algorithm first starts with extracting frequency adaptive attention weights from the time-pooled convolution input. This is achieved by attention module composed of two 1D convolution layers that squeeze the channel dimension into the number of basis kernel $K$, resulting in $K$ attention weights for all frequency bins. Then, weighted summation is applied to $K$ basis kernels with $K$ attention weights corresponding to each frequency bin. The basis kernels are also composed of trainable parameters, and the attention weights allow different combinations of basis kernels to generate an optimal kernel for different frequency bins.

FDY conv is proposed to release translational equivariance along the frequency dimension by using frequency-dependent convolution kernels [26]. However, this approach is not the only nor the simplest way to implement frequency-dependent convolution kernels. Therefore, we design simpler approaches. This is to implement frequency-dependent convolution kernels and compare their performance with FDY conv to gain a deeper understanding of the characteristics and effects of FDY conv on SED. The most straightforward method for implementing frequency-dependent convolution is to assign separate convolution kernels for each frequency bin, as illustrated in Fig. 3. (a). We named this approach as *frequency-wise kernel convolution (FK conv)*. This approach drastically increase the number of trainable parameters because it requires $F$ convolution kernels where $F$ refers to number of frequency bins of convolution input. A more complex, yet simpler method than FDY conv is to apply the weighted sum of trainable basis kernels using fixed frequency-dependent weights and as illustrated in Fig. 3. (b). This approach is named as *frequency-wise weighted convolution (FW conv)*. It is similar to FDY conv illustrated in Fig. 3. (c), with the only difference being that the weights are trained to be fixed for each frequency bin. Both FK conv and FW conv apply fixed convolution kernels assigned to a specific frequency bin. They apply different kernels on different frequency bins, resulting in the application of frequency-dependent kernels that release translation equivariance of 2D convolution on the frequency dimension.

## III. Experimental Setup

In this section, we provide details of SED model training framework used in this work. Basic format follows Detection and Classification of Acoustic Scenes and Events (DCASE) challenge 2022 task 4 baseline [49], [50]. Official code implementation of this work is publicly available in following GitHub repository: *https://github.com/frednam93/FDY-SED.*

### A. Implementation Details

The dataset used in this work is the Domestic Environment Sound Event Detection (DESED) dataset [49], consisting of audio recordings with a length of 10 seconds and a sampling frequency of 16 kHz. DESED is composed of real or synthetic audio clips those depict sound scenes commonly heard in domestic environments. The dataset includes ten sound event classes as shown in Table I, which also includes statistics of sound events in DESED real validation dataset. DESED contains a strongly labeled synthesized dataset, weakly labeled real dataset, and an unlabeled real dataset, as illustrated in the left side of Fig. 4. Strongly labeled datasets provide labels that include the sound event class within each audio clip, along with the event's onset and offset time within the audio. The weakly labeled dataset provides only the sound event class present in each audio clip, while the unlabeled dataset provides no labels. Synthesized strongly labeled datasets are created by combining foreground sound event clips with background sound clips to simulate soundscapes of domestic environments. The datasets consist of 10000, 2500, 1578, 14412 and 1168 audio clips for synthetic strongly labeled train dataset, synthetic strongly labeled validation dataset, real weakly labeled dataset, real unlabeled dataset, and real validation dataset respectively. Real weakly labeled dataset is randomly split into weakly labeled train dataset and weakly labeled validation dataset with a 9:1 ratio before training. During training, 12, 12, and 24 audio clips from the synthesized strongly labeled dataset, the weakly labeled train dataset, and the real unlabeled dataset, are batched respectively to form a batch consist of total 48 audio clips in every steps in each epoch. For validation, batch of 24 clips are randomly selected from the concatenated dataset of the synthesized strongly labeled validation dataset and the weakly labeled validation dataset. Finally, for test, the real validation dataset is used with batch size of 24.

TABLE I
Sound Event Classes Statistics in DESED Real Validation Dataset.

| Classes | Number of items | Average length (s) |
|---|---|---|
| Alarm/bell ringing | 420 | 1.96 |
| Blender | 96 | 5.13 |
| Cat | 341 | 1.38 |
| Dishes | 567 | 0.63 |
| Dog | 570 | 1.41 |
| Electric shaver/toothbrush | 65 | 7.73 |
| Frying | 94 | 8.26 |
| Running water | 237 | 5.25 |
| Speech | 1754 | 1.50 |
| Vacuum cleaner | 92 | 8.48 |

To leverage large amount of unlabeled dataset provided in DESED, semi-supervised learning framework is implemented using mean teacher method [49], [51]. We apply data augmentation on student model and teacher model with different random parameters as shown in the middle part of Fig. 4. Adam optimizer was used for 200 epochs. Single NVIDIA RTX Titan GPU is used for each training run.

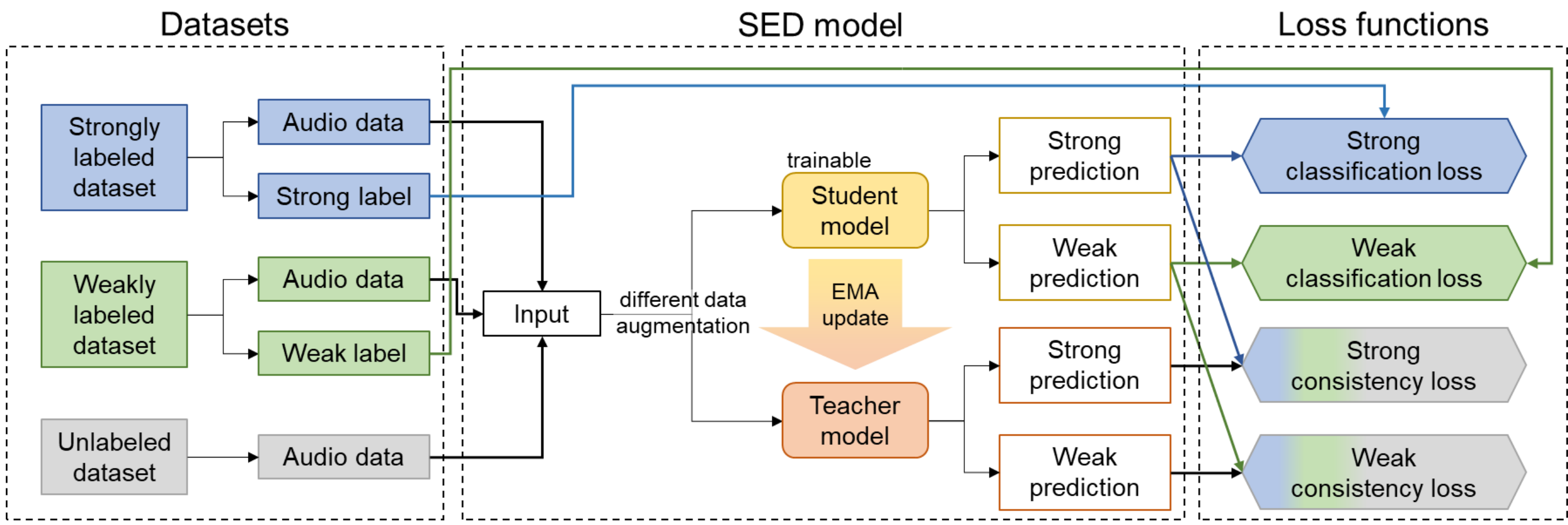

**Fig. 4.** An illustration of the framework for training the SED models in this work. It applies mean teacher algorithm with strongly labeled dataset, weakly labeled dataset and unlabeled dataset to minimize four loss functions: strong classification loss, weak classification loss, strong consistency loss and weak consistency loss.

*B. Input Feature*

We first normalize the audio clips' waveforms to set the maximum of absolute value to be one, and then we extracted mel spectrograms from it. To extract mel spectrograms, short time Fourier transform (STFT) is applied to the waveform with number of fft, hop length of 2048 and 256 respectively. Hamming window is used to window within STFT. Then mel scale with 128 mel bins is applied to resultant spectrogram to obtain mel spectrogram. Though the y axis of the input feature represents mel dimension, we refer to it as frequency dimension in this work for explanation consistent to the motivation of this work. The numbers we used followed standard configurations used in prior SED literature and DCASE Challenge baselines for all audio feature extraction parameters [52].

*C. Data Augmentation*

Several data augmentation methods are applied in this work. Frameshift [49], mixup [53] and time masking [13] are applied as default as in [5], [26], [35]. These methods involves change in labels as well. Mixup is applied on both strongly labeled data and weakly labeled data. Time masking is applied to the audio data and the corresponding label at the same time to make sure the label also implies no event exist on the time duration where input mel spectrogram is masked. Then we apply one of the frequency-wise data augmentations for Grad-CAM evaluation. The optimal hyperparameters follow the previous work [35]. For analysis on frequency dependent convolutions, we do not apply any of frequency-wise data augmentation. For overall & performance analysis and PCA analysis on FDY-CRNN, we apply optimal setting of step type FilterAugment from [26].

*D. Baseline Model Architecture*

The baseline SED models used in this work are based on a CRNN architecture with an attention module for weak prediction training [49]. The module consists of a CNN, an RNN, and a fully connected (FC) layer with an attention mechanism as described in Fig. 5. The CNN has seven convolutional layers, each composed of a 2D convolution module followed by 2D batch normalization, an activation, dropout with a probability of 0.5, and then 2D average pooling. Note that when we apply convolution modules with frequency-dependent kernels, we replace 2$^{nd}$ ~7$^{th}$ convolution layers as in [26]. Activation is rectified linear unit (ReLU) for Grad-CAM analysis and context gating (CG) for the rest [54]. The pooling modules reduce the frequency dimension by a factor of two in all convolutional layers, while they only reduce the temporal dimension by a factor of two in the first two convolutional layers. The output of the CNN then enters the RNN, which consists of two bidirectional gated recurrent unit (biGRU) modules followed by dropout with a probability of 0.5. Finally, single FC layer followed by a sigmoid function squeeze the channel dimension of the RNN output into the class dimension in a time-wise manner, resulting in strong prediction which include the time span of the predicted sound event class. After the fully connected layer, the attention module pools the representation along the time axis to obtain weak predictions, which only indicate the presence of target sound events without temporal information.

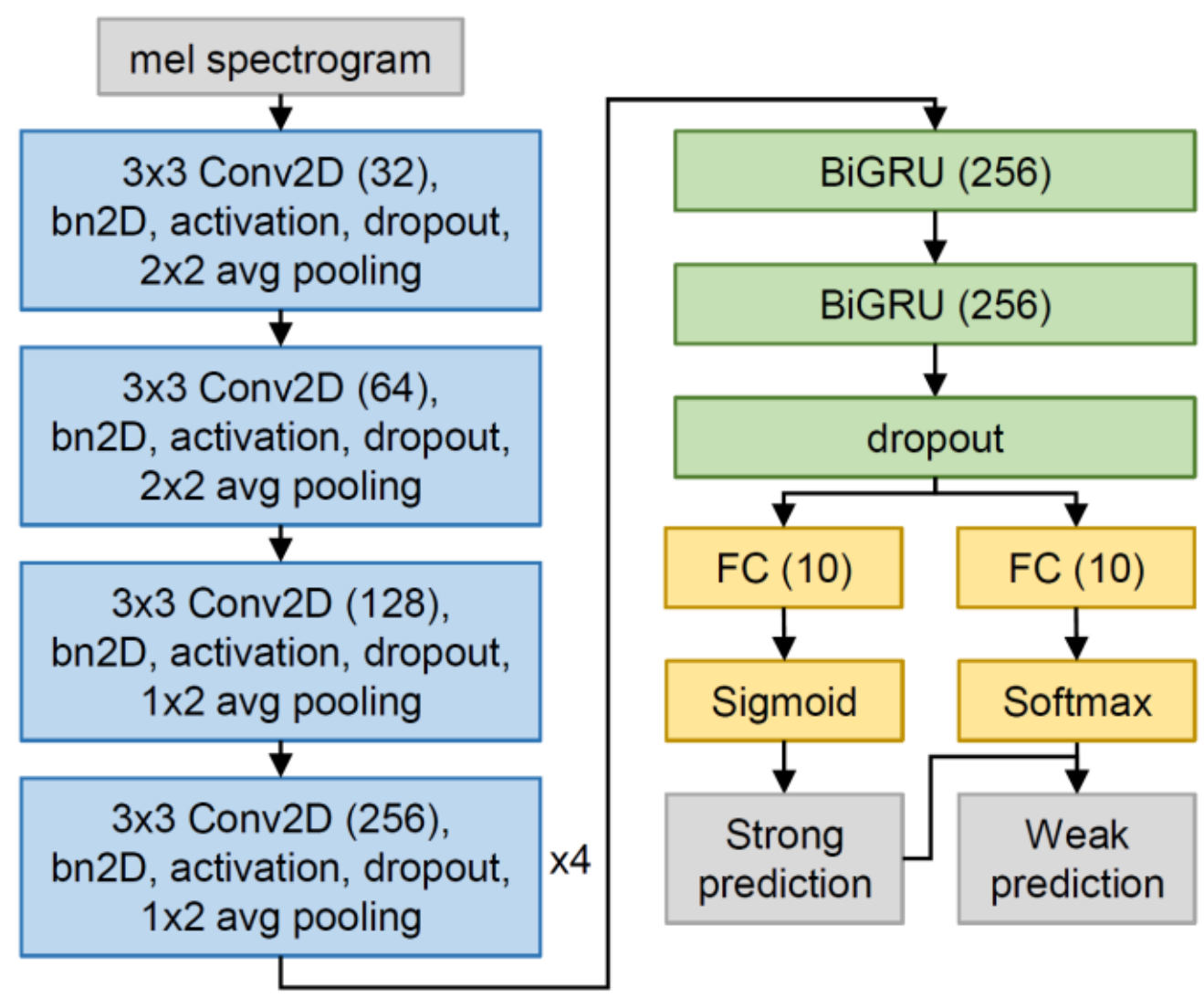

**Fig. 5.** An illustration of SED baseline model architecture used in this work.

TABLE II
ABLATION STUDY ON FILTER SIZE AND NUMBER OF CONVOLUTIONAL LAYERS IN THE BASELINE SED MODEL.

| Settings | CB-F1 | PSDS1 | PSDS2 |
|---|---|---|---|
| baseline | 0.478 | 0.3957 | 0.5983 |
| Half filter size | 0.465 | 0.3891 | 0.5924 |
| Double filter size | 0.488 | 0.3937 | 0.6187 |
| 8 conv layers | 0.481 | 0.4051 | 0.6394 |

To validate the chosen architecture, we conducted an ablation study on the number of convolutional layers and the filter size. As summarized in Table II, doubling the filter size or increasing the number of layers to 8 led to moderate performance improvements. However, the baseline setting was selected as a trade-off between performance and complexity, as the primary focus of this work is on analyzing frequency-dependent modeling rather than optimizing model capacity.

### *E. Computational Complexity*

To analyze the computational cost of FDY Conv, we provide a complexity comparison against standard 2D convolution. FDY Conv utilizes a set of K basis convolution kernels and computes frequency-dependent attention weights to dynamically combine the outputs at each frequency bin. The overall computational complexity can be decomposed into three components. First, the attention weights are calculated through a two-layer fully connected network applied over the frequency axis. We follow the common practice of reducing the channel dimension by a factor of $r$ in the intermediate layer, resulting in a complexity of $O(F \cdot (C_{in}^2/r + CK))$, where $F$ is the number of frequency bins, $C_{\text{in}}$ is the number of input channels, and $K$ is the number of basis kernels.

Second, each of the $K$ basis kernels performs a full 2D convolution over the input feature map. Assuming a kernel size of $K_{\text{f}} \times k_t$ over the frequency and time axes, the total convolution cost becomes $O(K \cdot B \cdot C_{in} \cdot C_{out} \cdot K_f \cdot K_t \cdot F \cdot T)$, where $B$ is the batch size, $C_{out}$ is the number of output channels, and $T$ is the number of time frames.

Third, the outputs of the $K$ basis kernels are combined via weighted summation using the attention weights for each frequency bin, which incurs an additional cost of $O(B \cdot C_{out} \cdot F \cdot T \cdot K)$.

In contrast, a standard 2D convolution layer has a single kernel and a complexity of $O(B \cdot C_{in} \cdot C_{out} \cdot K_f \cdot K_t \cdot F \cdot T)$. Although FDY Conv introduces additional overhead due to the attention mechanism and multiple basis kernels, the increased cost is proportional to K and remains manageable in practice. The performance improvements shown in our experiments suggest that this trade-off is worthwhile.

TABLE III
PERFORMANCE COMPARISON OF FREQUENCY DEPENDENT METHODS ON SED.

| Settings | CB-F1 | PSDS1 | PSDS2 | Improvement |
|---|---|---|---|---|
| baseline | 0.478 | 0.3957 | 0.5983 | |
| +FreqMask | 0.474 | 0.4036 | 0.6204 | +3.02% |
| +FiltAug | 0.486 | 0.4097 | 0.6337 | +4.97% |
| +FDY | 0.505 | 0.4337 | 0.6493 | +8.95% |
| +FDY +FM | 0.493 | 0.4386 | 0.6595 | +10.47% |
| +FDY +FA | **0.517** | **0.4410** | **0.6683** | **+11.60%** |

### *F. Loss Function*

As shown in the right side of Fig. 4, the loss function in this work consists of four terms [49]: strong classification loss, weak classification loss, strong consistency loss, and weak consistency loss, as in [49]. The loss can be represented by following equation:

$$L = B(SP_s, l_s) + w_W \times B(WP_s, l_w) + w_c \times L_{cons} \quad (2)$$

where $B(x, y)$ refers to binary cross-entropy (BCE) loss between $x$ and $y$. $w_W$ and $w_c$ refer to weight for weak classification loss and consistency losses. $SP_s$ and $WP_s$ refer to strong and weak predictions by student model respectively. $l_s$, and $l_w$ refer to strong and weak labels respectively. $L_C$ refers to consistency loss expressed by:

$$L_{cons} = M\big(SP_s, sg(SP_T)\big) + M\big(WP_s, sg(WP_T)\big) \quad (3)$$

where $M(x, y)$ refers to mean square error (MSE) loss between $x$ and $y$. $sg(x)$ refers to stop gradient operation on $x$. $SP_T$ and $WP_T$ refers to strong prediction and weak prediction by teacher model respectively. For polyphonic SED where multi-label cases exist, BCE loss is used for both strong and weak classification losses. The strong classification loss is applied only to strongly labeled data within the batch, while the weak classification loss is applied only to weakly labeled data within the batch. $w_W$ of 0.5 is used. Consistency loss is applied for semi-supervised learning, to ensure that the student model's predictions are consistent with the teacher model's predictions. Consistency loss is applied to the whole batch, including strongly labeled data, weakly labeled data, and unlabeled data using MSE loss. $w_c$ exponentially increases from zero to two at every step within each epoch for the first fifty epochs.

### *G. Post Processing*

After obtaining predictions, we perform weak prediction masking on strong predictions by applying the weak predictions of the audio data as a class-wise threshold for the strong data prediction [5]. Specifically, we mask the strong prediction values of each sound event class as zero on the time frames where the values are below the weak prediction values for that sound event class. This is done because training weak predictions is more robust than training strong predictions. Strong predictions suffer from undertraining due to the abundance of inactive frames in strong labels, while weak predictions do not. Additionally, DESED dataset uses a synthesized dataset for strongly labeled data while using a real dataset for weakly labeled data. After weak prediction masking, we apply a median filter of length 7 as set in DCASE challenge baseline which optimized size of class-independent median filter [49], which corresponds to approximately 0.45 seconds in time. Unlike previous work that applied class-wise median filters, we used a simpler setting to focus on the effect of other frequency-dependent methods [26]. As a result, the performances in Tables II and IV are worse than in [26], [35].

TABLE IV
PERFORMANCE OF MODELS IN GRAD-CAM ANALYSIS.

| model | CB-F1 | PSDS1 | PSDS2 | LM - GC |
|---|---|---|---|---|
| no fw aug | 0.412 | 0.3445 | 0.5761 | 0.2244 |
| freq mask | 0.423 | 0.3391 | 0.5427 | 0.2222 |
| step FiltAug | 0.436 | 0.3452 | 0.5703 | 0.2221 |
| lin FiltAug | **0.440** | **0.3559** | **0.5805** | **0.2212** |

### *H. Evaluation Metrics*

To evaluate the performance of various SED methods, we employ two metrics: collar-based F1 score and polyphonic sound detection score (PSDS). The collar-based F1 score (CB-F1) compares the onset and offset of each sound event class in both the label and SED predictions [3]. A prediction is considered a true positive if its onset and/or offset falls within the collars of the label's onset and/or offset. This metric can also provide class-wise scores. The PSDS metric addresses some limitations of the CB-F1 score by considering the intersection between the prediction and label, as well as cross triggers that are induced by other sound events present in the audio [55]. PSDS also employs receiver operating characteristic (ROC) curves, which allows for comparison of SED performances without requiring threshold optimization. Two types of PSDS (PSDS1 and PSDS2) are used in the DCASE Challenge 2021 and 2022 Task 4 to evaluate SED systems [49]. PSDS1 emphasizes accurate time localization by limiting tolerance on the intersection criteria, while PSDS2 emphasizes accurate classification by penalizing cross triggers more.

In this work, we compare the SED performances of various settings using PSDS1 + PSDS2, as used to rank entries for DCASE Challenge 2021 and 2022 Task 4. We run 12 separate trainings to obtain 24 models (including both student and teacher models) and reported the maximum score achieved in the tables. Additionally, we report the macro collar-based F1 score for further information. As PSDS does not support class-wise scores, we use CB-F1 score to compare the performance of each sound event class.

## IV. RESULTS AND ANALYSIS

### *A. Performance comparison*

The comparison of SED models presented in Table III indicates the effectiveness of FilterAugment in comparison with frequency masking (FreqMask) and FDY conv in improving the performance of SED. The results suggest that the application of FilterAugment with FDY conv in combination leads to an improvement of 11.6% over the baseline. In particular, if we assume that these two methods are completely unrelated, we would expect the improvement obtained by their joint application to be equal to the product of their individual improvements, which would be 14%. Considering that the observed improvement of 11.6%, the result suggests that FilterAugment and FDY conv are mostly unrelated. Taken together, these findings suggest that while both FilterAugment and FDY conv both address frequency-related issues in SED, their roles are complementary rather than overlapping. This highlights the importance of considering multiple techniques in the development of SED models, and the potential for achieving further improvements through the combination of different methods. In addition, FilterAugment outperforms frequency masking with and without FDY conv.

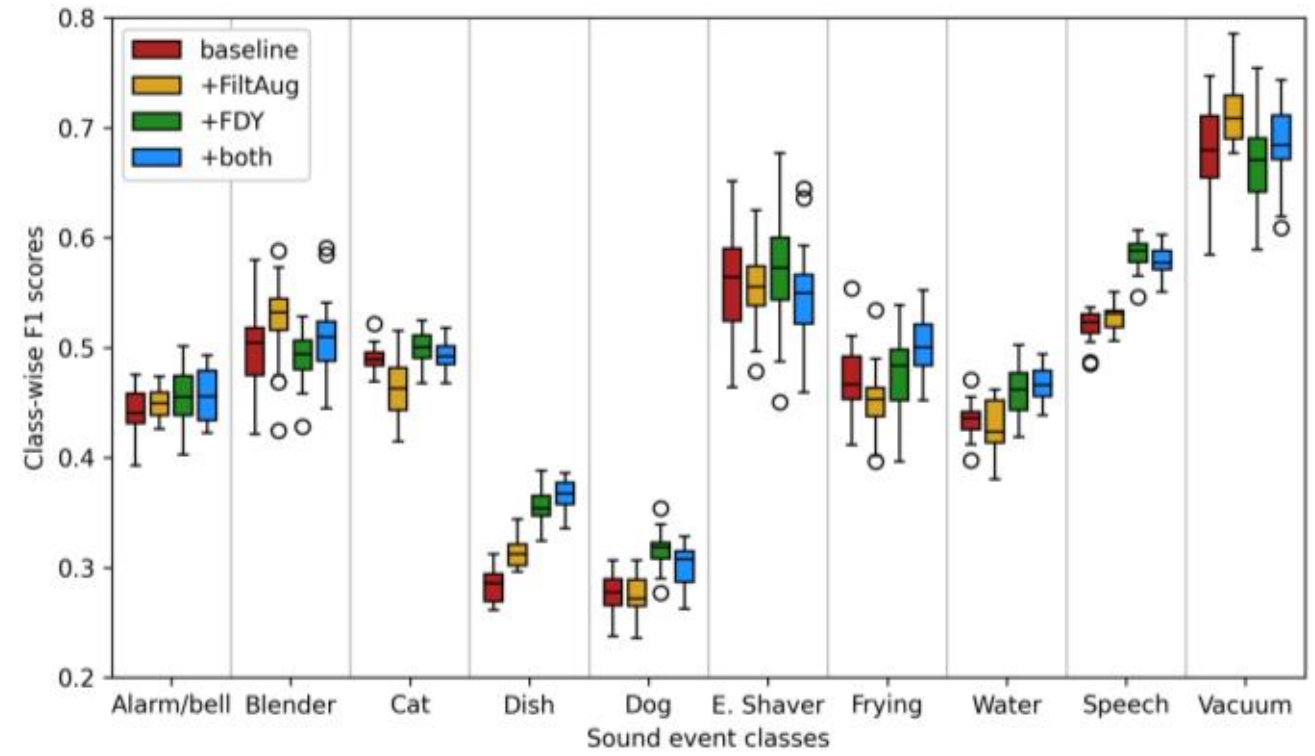

**Fig. 6.** Boxplots for class-wise comparison between baseline model, model with FilterAugment, model with FDY conv, and the model with both.

To gain deeper insights into the performance of these methods on SED, we conducted a class-wise performance comparison of four SED model settings. We used the CB-F1 score to provide class-wise performance and plotted box plots of CB-F1 scores for 24 models, including student and teacher models, for each model setting on Fig. 6. Our analysis revealed that in general, the performance is improved in the order of baseline, with FilterAugment, with FDY conv and with both, similarly to the overall performance results in Table III. These tendencies are particularly evident for sound events such as alarm/bell ringing, dishes, and running water.

However, we also observed several exceptions to this general trend. For instance, adding FDY conv resulted in worse performance for sound events such as blender and vacuum cleaner, which are stationary sound events. Stationary sound events are those almost do not change in their frequency contents over time. Sound events such as blender, electric shaver/toothbrush, frying, running water and vacuum cleaner involve stationary sound components. On the other hand, FDY conv is more specialized on detecting complex non-stationary sound patterns by applying frequency-adaptive kernels. Unlike the conclusion in the previous work [26], FDY conv improved the performance for sound events having stationary sound components such as electric shaver/toothbrush, frying, and running water. It is because, on the top of stationary components, these three sound events involve partially non-stationary components. Electric shaver/toothbrush often involves impulsive sound when cutting stiff beard, frying involves frequent impulsive popping sound, and running water involves human interaction upon use of the water in domestic environment, causing non-stationary sound. Thus, by comparing the class-wise performance of the baseline and FDY conv, we can conclude that FDY conv is especially strong on non-stationary sound events.

We also discovered that adding FilterAugment caused worse performances for sound events such as cat, dog, and electric shaver/toothbrush. In case of frying, FilterAugment only worsened the performance on baseline, but not on the model with FDY conv. These sound events are characterized by specific and local frequency components. The sounds produced by cats and dogs are mainly composed of meowing and barking, respectively. These sounds are characterized by the frequency of the animals' vibrating vocal cords, which results in specific and localized frequency peaks. Electric shavers and toothbrushes contain rotating motors that produce a steady vibrating sound with few prominent frequencies. In case of frying, its sound does not involve a single localized frequency peak, but consists of locally banded random noise in the high-frequency range. With regard to the other sound events, FilterAugment shows a consistent improvement in performance. This is because FilterAugment is able to extract sound event information from a wider frequency range, making it less effective for sound events that have specific and localized frequency peaks. Our analysis suggests that FilterAugment extracts sound event information from a wider frequency range and may not be suitable for sound events with specific and localized frequency peaks.

This study demonstrates that both FilterAugment and FDY conv significantly improve the performance of SED models. However, the effectiveness of these methods varies across different types of sound events, and their performance may be influenced by the specific acoustic characteristics of the events in question. While our method is not intended to compete with recent transformer-based SOTA models, it offers architectural insights and interpretable analyses on frequency-adaptive CNN structures, many of which form the basis of hybrid SED systems. Our findings are thus complementary to performance-driven models and highlight design choices applicable across architectures.

### *B. Grad-CAM on FilterAugment*

Grad-CAM is a visualization method that highlights input regions contributing most to a specific class prediction in convolutional neural networks (CNNs) by back-propagating gradients from the output to intermediate feature maps [21], [47]. However, Grad-CAM is proposed for CV methods thus the algorithm required several modifications for SED. While image recognition yields only one prediction per one image, SED yields predictions corresponding to several time frames of input audio. Thus we expanded Grad-CAM's target tensor dimension to add the time dimension. In addition, since SED allows multi-label situations where multiple sound events can occur simultaneously and overlap on the spectrogram in the same time-frequency region, we set the target of audio as each sound event class and we focused the analysis on the classes that are present on both the label and predicted class. We applied Grad-CAM to the entire CNN module of the SED model, from the first to the last convolutional block—including convolution, batch normalization, ReLU, and pooling layers—to capture the overall behavior of the CNN in response to frequency-wise data augmentations. By specifying the target layers, Grad-CAM automatically generates attention maps that match the spatial resolution of the corresponding feature maps. These maps were plotted in Fig. 7 to visualize the model's focus under different training conditions. Following the standard Grad-CAM formulation [47], the attention map for sound event class $c$ is computed as a weighted combination of the feature maps $A^k$ from the target laeyrs:

$$L^c_{Grad-CAM} = ReLU(\sum_k \alpha^c_k A^k) \quad (4)$$

where $\alpha^c_k$ denotes the average gradient of the tatget class score $y^c$ with respect to the feature map $A^k$, given by:

$$\alpha^c_k = \frac{1}{Z} \sum_i \sum_j \partial y^c / \partial A^k_{ij} \quad (5)$$

where $Z$ denotes number of pixels of feature map.

For this analysis, slightly simpler model architecture is used to isolate the effects of frequency masking and FilterAugment. We used conventional 2D convolution modules and used ReLU for activation, because CG is a complex activation function with learnable parameters. SED performances of the models used in Grad-CAM analysis are presented in Table IV. It can be observed that both types of FilterAugment perform better than model without frequency-wise data augmentation and model with frequency masking in terms of CB-F1, PSDS1 and PSDS2. Also, linear type FilterAugment performs better than step type FilterAugment as in [35]. The Grad-CAM plots shown in Fig. 7. demonstrate the effects of these methods on example audio data having following sound events classes: (a) cat, (b) dish, and (c) frying. The amplitude of the Grad-CAM plots indicates that the SED models have extracted sound event class information from the time-frequency regions with larger values (closer to one). Furthermore, we visualize the difference between each model's Grad-CAM and the input log-mel spectrogram using subtraction maps, denoted as "LM – GC" in Fig. 7. These plots display the squared difference between the log-mel and Grad-CAM outputs, averaged over time-frequency regions where sound events are active. This provides a quantitative measure of how closely the model's attention aligns with the original input features. In addition, Table IV presents the LM–GC values for each model, computed over the entire test set. Although the differences are subtle, we observe a consistent decrease in LM–GC from no augmentation to frequency masking, step FilterAugment, and linear FilterAugment. While LM–GC provides a useful indication of the alignment between model attention and input features, it should not be interpreted as a direct proxy for model performance. A lower LM–GC value suggests that the attention map is more input-consistent, but this does not necessarily translate to higher detection accuracy. Therefore, we use LM–GC as a complementary interpretability metric rather than a primary evaluation criterion.

Our observations reveal that FilterAugment effectively helps SED models to accurately detect time-frequency regions closely associated with sound events. For instance, from Fig. 7.

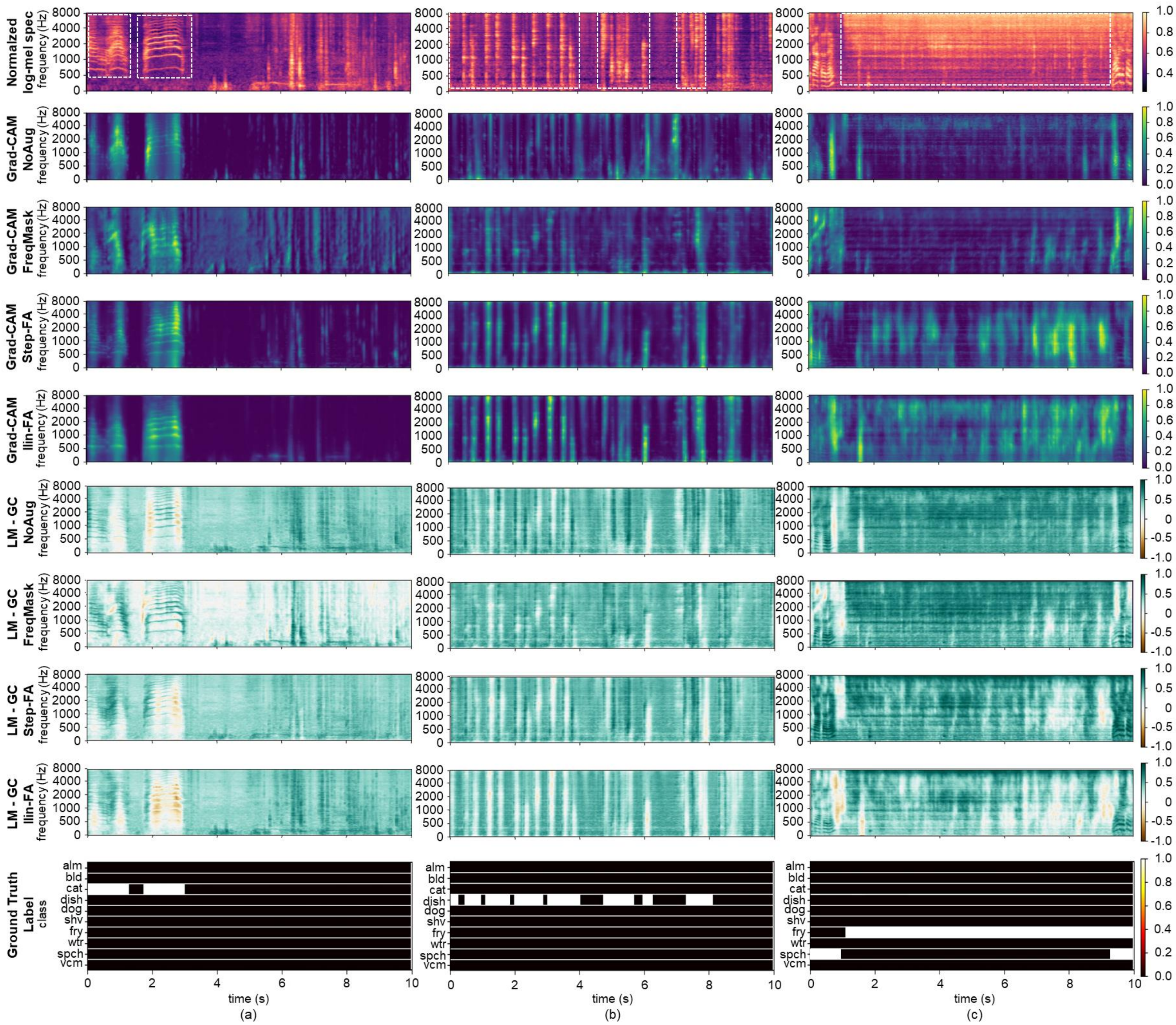

**Fig. 7.** Examples of Grad-CAM applied on SED models. The plots illustrate the normalized log-mel spectrogram of the input audio, Grad-CAM on SED models without frequency-wise data augmentation, with frequency masking, with step FilterAugment and linear FilterAugment, subtracted each Grad-CAM from log-mel spectrograms and the ground truth label corresponding to the data. The plots depict audio data on sound event classes of (a) cat (Y0R4uO4Vy9c0_110.000_120.000.wav), (b) dish (Y1QK5WA4dK4o_400.000_410.000.wav) and (c) frying (Y1U4D63urKlo_90.000_100.000.wav).

(a), we can observe that SED models with frequency masking extract sound information from other time durations that are unrelated to the sound event by comparing Grad-CAM plot and ground truth label plot. While Cat sound is active on 0 ~ 3 seconds in ground truth label plot, Grad-CAM plot for frequency masking is active over entire time duration consistently showing value over 0.5 for whole time ranges. On the other hand, SED model without frequency-wise data augmentation extract negligible amount (below 0.5) of information from those unrelated durations on 3 ~ 10 seconds. This is consistent to the claim in [35] that frequency masking could make SED model to forcibly extract sound event information from unrelated time-frequency regions. We can infer that SED models with FilterAugment can effectively infer the formants of the cat's voice as in the log mel spectrogram plot in the white dashed box around on 2 ~ 3 seconds. The formants are not shown in Grad-CAM by model without augmentation and it is vaguely shown in Grad-CAM by model with frequency masking. This can also be observed in the subtracted plots of Grad-CAM from log-mel spectrogram, that the formant area has green color meaning high value for Grad-CAM with no augmentation or frequency masking, while it has white color meaning small difference on Grad-CAM with FilterAugment. Since the audio includes meowing sound of cat, FilterAugment effectively extract cat sound information from the details such as formants. From this observation, we can infer that step type FilterAugment focus on local and characteristic features while linear type FilterAugment extract information from less confined range of features in terms of frequency components.

TABLE IV
PERFORMANCE OF MODELS USING CONVOLUTION METHODS WITH FREQUENCY-DEPENDENT KERNELS.

| Settings | CB-F1 | PSDS1 | PSDS2 | Improvement |
|---|---|---|---|---|
| baseline | 0.478 | 0.3957 | 0.5983 | |
| +FK conv | 0.430 | 0.3367 | 0.5116 | -14.66% |
| +FW conv | 0.490 | 0.4078 | 0.6232 | +3.72% |
| +FDY conv | **0.505** | **0.4337** | **0.6493** | **+8.95%** |

From Fig. 7. (b), by observing the color of vertical streaks, we can observe that SED models with FilterAugment recognize short impulsive dish sounds from the mel spectrogram better as their Grad-CAM plots show brighter vertical streaks, showing values of 0.8~1.0. On the other hand, Grad-CAM plots by SED models with no augmentation or frequency masking show darker streaks with values around 0.4 thus they inferred less from the the short impulsive sound of dishes. This can be observed in subtracted plots, as plots for Grad-CAM with FilterAugment shows more white vertical streaks. We can observe that Grad-CAM plot by SED model without frequency-wise data augmentations highlights two impulsive sounds at 6 second and 7 second, while the ground truth label plot shows that the one at 7 second is a false target. SED model with frequency masking is better but its Grad-CAM do not highlight all the vertical streaks corresponding to labels compared to those by FilterAugment. On the other hand, SED model with both step and linear type FilterAugment highlights impulsive sounds corresponding to the true dish sounds illustrated by ground truth label. When comparing Grad-CAM by step type and linear type FilterAugment, we can also observe that the SED model with step type FilterAugment accentuates three to five very specific frequency bins within the vertical streaks which makes the streaks look like bamboo with joints. On the other hand, SED model with linear FilterAugment accentuates rather wider ranges of frequency within the vertical streak. While two types of FilterAugment show similar performance gain on SED, they give different specialties to SED models; step type FilterAugment makes SED models to focus on narrow frequency region while linear type FilterAugment makes SED models to focus on rather wider frequency region.

Lastly, from Fig. 7. (c), SED models without augmentation or with frequency masking barely inferred sound information from the time range from 1 second to 10 second, where the events exist. However, SED models with FilterAugment can extract sound information from this region more effectively. Notably, step type FilterAugment focuses more on the impulsive popping sounds ocassionally happening during frying, which are observed at 1.5, 7, 7.5, 8 and 9 seconds as vertical streaks. On the other hand, linear FilterAugment focuses widely on the broad-banded white noise which are temporarily stationary as well. The subtracted plots also shows more white parts for Grad-CAM with FilterAugment, implying more similarity between log-mel spectrogram and Grad-CAM for models with FilterAugment.

We emphasize that the Grad-CAM results in Fig. 7 are presented as case studies to qualitatively interpret model behavior under different training conditions, rather than to statistically validate performance differences. While multiple examples can help highlight typical patterns, the inherent variability across instances limits their generalizability. These examples are therefore intended to illustrate observable trends in model attention, not to claim consistent superiority across all samples.

The above discussions have shown that FilterAugment helps SED models to effectively recognize time-frequency regions closely associated with sound events. Furthermore, while step type FilterAugment makes SED model to focus on more local and sharply distinct time-frequency region to infer the sound event information, linear type FilterAugment makes SED model to focus on less local and wider range of time-frequency region. This is also consistent to their algorithm where step type FilterAugment results in amplitude changed discretely over frequency dimension while linear type FilterAugment results in amplitude changed continuously over frequency dimension. More interesting discovery is that while two types of FilterAugment alter amplitude over only frequency dimension, they focus on discrete or continuous feature along not only frequency dimension but also time dimension.

### *C. Experiments on Convolution Methods with Frequency-Dependent Kernels*

We evaluated the effectiveness of three convolution methods with frequency-dependent kernels on the baseline SED model without FilterAugment. We listed the performance of each method in Table IV.

The results show that FK conv deteriorated the performance (-14.66%), while FW conv slightly enhanced (+3.72%), and FDY conv greatly improved the baseline model (+8.95%). Comparing the performances between FK conv and FW conv, we can infer that using separate kernels on each frequency bin is simply inefficient since the performance deteriorated in table IV. Although the time-frequency patterns are dependent on frequency, they share similarities as well. Adjacent frequency bins exhibit some similarities as we can still recognize sound information when audio is applied with a slight pitch shift. In such context, FW conv outperformed the baseline in table IV by applying the idea of frequency-dependent kernels with shared basis kernels. All frequency bins use kernels that are linear combinations of basis kernels, which could learn the shared sound patterns across the frequency dimension.

In FDY conv, the attention module infers the frequency contents from the convolution input and applies an appropriate frequency-dependent kernel to effectively extract sound information, outperforming the baseline by 8.95%. This does not merely claim the excellence of attention mechanism which adapts to the contents of convolution input. This shows that attention mechanism has well fitted to FW conv by providing weights that adapt to frequency contents by convolution input. In addition, considering that acoustic waves follow the principle of superposition, meaning that simultaneous sound events and acoustic scenes could be described as superposition of these sounds in waveform domain, using weighted sum of basis kernels could be explained as recognizing various superposed sound event patterns. This explains the synergy between the attention mechanism and FW conv: the attention mechanism helps to identify which pattern should be highlighted more given the data, then FW conv extract sound information corresponding to those patterns and then superpose the extracted information accordingly.

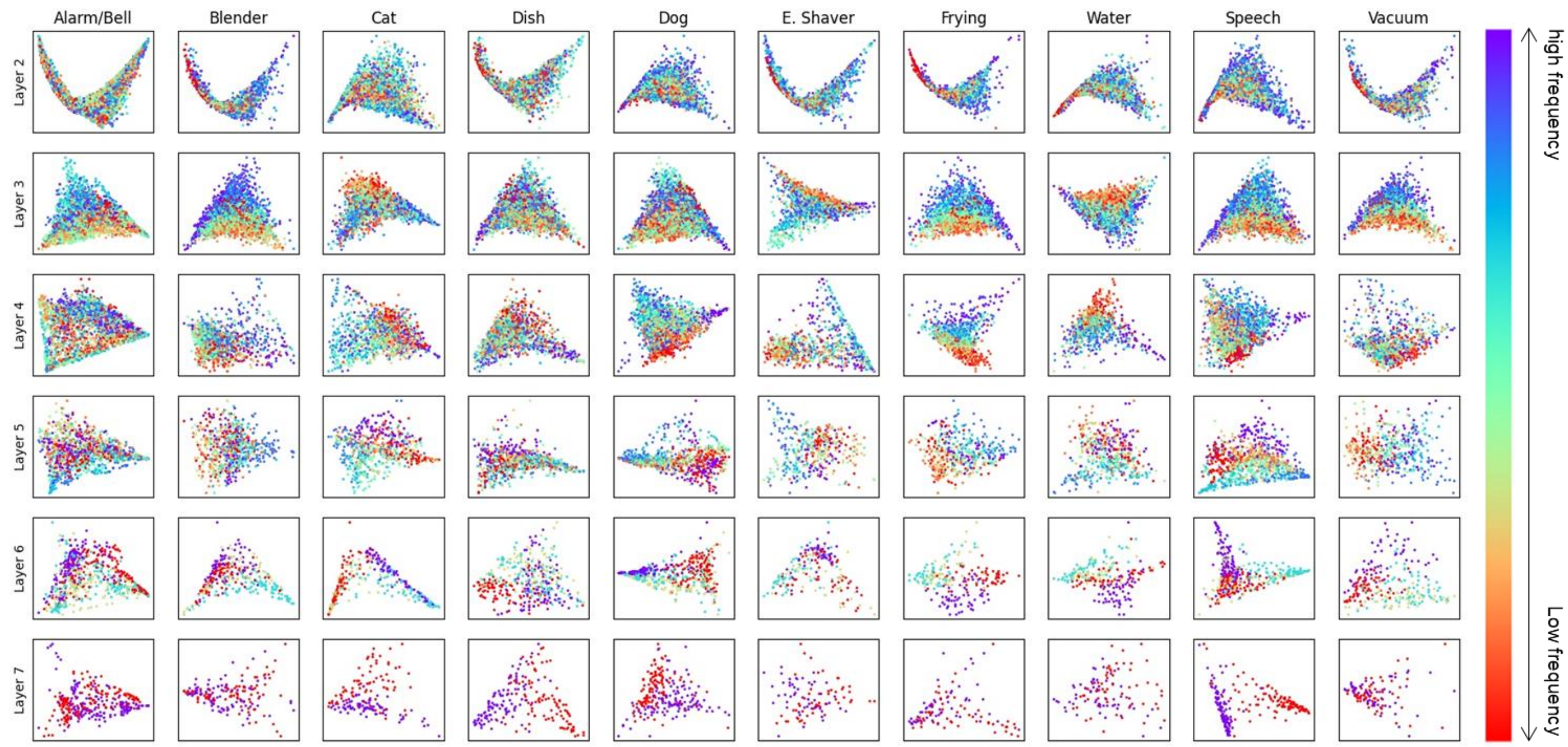

**Fig. 8.** Examples of PCA analysis on the attention weights used in frequency dynamic convolution.

### *D. PCA analysis on the attention weights from Frequency dynamic convolution*

FDY conv uses convolution kernels that vary along frequency bin, applying weighted summation on the basis kernels with attention weights. To observe how attention weights change across frequency dimension within different sound event classes, we applied principal component analysis (PCA) to the attention weights [20]. We used PCA to extract the two most significant dimensions to obtain 2D plots which present concise information. To obtain frequency-adaptive attention weights corresponding to each sound event class, we prepared audio clips containing single sound event class spanning 10 seconds using foreground dataset in DESED which is the material of synthesized strongly labeled datasets [49]. The foreground dataset is composed of audio clips those only having foreground sound events corresponding to one of ten sound event classes. Applying FDY-CRNN on the foreground dataset, we obtained attention weights for each frequency bin of each audio data containing a single sound event class. We then performed PCA on the attention weights corresponding to each sound event class and plotted the two principal components with color representing the frequency to observe how the components change over the frequency bins within the sound event class.

Fig. 8. displays PCA analysis plots, representing two most significant component axises, on the frequency-adaptive attention weights within FDY conv. Each plot is composed of first two principal components of attention weights in x and y axis. Each column corresponds to a specific sound event class, and each row corresponds to a specific FDY conv layer. The frequency bin of dots in the plots are expressed using spectrum color ranging from red to purple, from low frequency to high frequency respectively.

The distributions shown in the PCA plots represent the structural behavior of frequency-adaptive attention weights across frequency bins within each sound event class. When frequency bins (color-coded) form local clusters, it implies that the model applies similar attention to neighboring frequencies, indicating semantically consistent adaptation. The distinct distributions across classes suggest that FDY conv dynamically adjusts frequency emphasis depending on the target class, supporting its effectiveness in learning class-specific spectral representations. As a result, we can infer that FDY conv applies convolution kernels those are optimal for each frequency contents of different sound event classes. If the attention weights did not depend on the frequency, the PCA plots would be rather random, making it difficult to observe the flow of spectral pattern within the plots.

More specifically, we can observe a clearer distinction between the colors in later layers, indicating that frequency-dependent attention weights are more distinctive across frequency in later layers. The plots of layer 2 show dots with various colors everywhere, though it exhibits some vague color patterns. However, from layer 3, similar colors are closer to each other.

The results highlight the importance of frequency-dependent attention weights in identifying different sound event classes. We can observe spectrally flowing patterns and dots with similar color cluster together, implying frequency-adaptive attention weights produce frequency-dependent convolution kernels.

We note that this PCA analysis was performed on the foreground dataset, which contains clean audio clips with single sound event classes. This setting was intentionally chosen to isolate and highlight the frequency-adaptive behavior of FDY conv without interference from overlapping events or background noise. While the observed class-wise separability may be less distinct in real monophonic or polyphonic mixtures, this controlled analysis offers foundational insights into how FDY conv attends to different frequency patterns across classes. Future work may extend this analysis to real-world datasets with more complex event structures to further validate these findings.

## V. Conclusion

This work aimed to explore the effectiveness of the frequency-dependent SED methods by various analysis methods. Class-wise performance comparison has shown that FilterAugment is effective on sound events of which energy is distributed over wide frequency range and FDY conv is superior in detecting non-stationary sound events. Grad-CAM analysis has proved that that SED models with FilterAugment infers sound event information from time-frequency regions closely related to the target event class. Further, step type FilterAugment focuses more on local time-frequency patterns while linear type FilterAugment focuses on larger and continuous patterns. Comparison of FDY conv with FK conv and FW conv has shown that while frequency dependent kernel only benefits SED when it considers shared pattern similarities among close frequency bins. Then we also showed that adding attention module further optimizes frequency adaptive kernels to contents in the frequency bins. PCA analysis has proved that kernels apply similar kernel on neighboring frequency bins within the same sound events, and kernels gradually changes over frequency dimension inferring that FDY conv applies frequency-dependent convolution kernels. We not only proved the superior performance by FilterAugment and FDY conv those outperform the baseline model by 11.6% when applied together, we also further broadened insights regarding these methods by delving into detailed characteristics and behaviors.

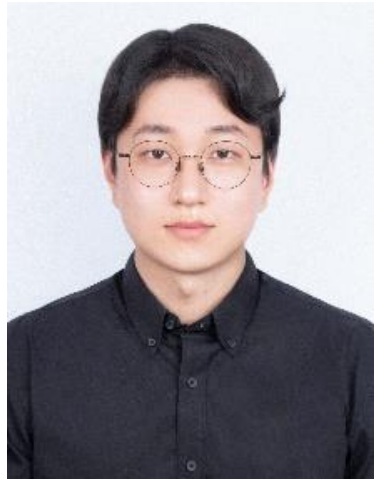

**Hyeonuk Nam** received the B.S., M.S., and Ph.D. degrees in Mechanical Engineering from the Korea Advanced Institute of Science and Technology (KAIST), Daejeon, Korea, in 2018, 2020, and 2025, respectively. He is currently with the Mobile eXperience Division, Samsung Electronics.

His research interests include various topics in auditory intelligence, such as sound event detection, automatic audio captioning, and text-to-audio generation. He is also interested in large language models (LLMs) and multimodal AI involving auditory, visual, and language modalities.

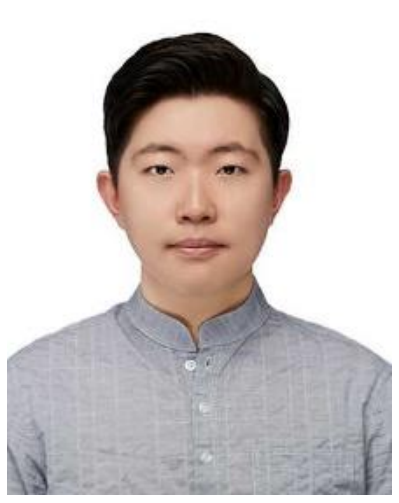

**Seong-Hu Kim** received B.S., M.S. and phD in mechanical engineering from Korea Advanced Institute of Science and Technology, Daejeon, Korea, in 2017, 2019 and 2024 respectively. He is currently working in Samsung Research.

His research interest includes text-independent speaker identification, text-independent speaker verification, and sound event detection.

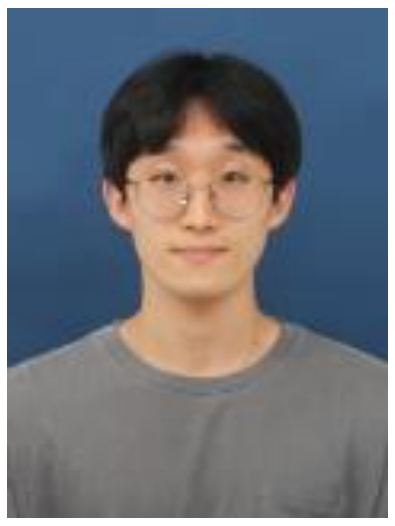

**Deokki Min** received B.S. degree in mechanical engineering from Yonsei University, Seoul, Korea, in 2022, and ths M.S. degree in mechanical engineering from KAIST, Daejeon, Korea.

His research interest includes sound event detection, acoustic representation, human hearing and auditory neuroscience.

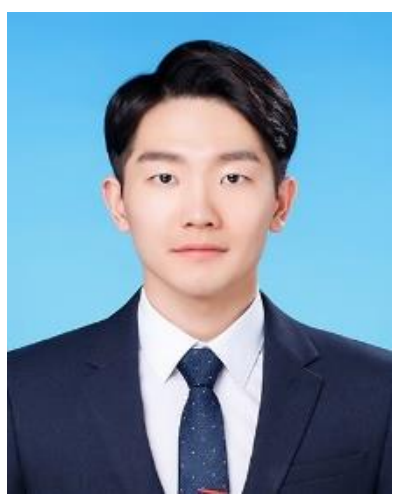

**BYEONG-YUN KO** received the B.S. degree in ship architecture and ocean engineering from Inha University, South Korea, in 2019, and the M.S. degree in mechanical engineering from KAIST, Daejeon, South Korea, in 2021. His research interest includes the development of deep learning-based individualization of HRTF and spatial audio rendering technique considering the human auditory perception.

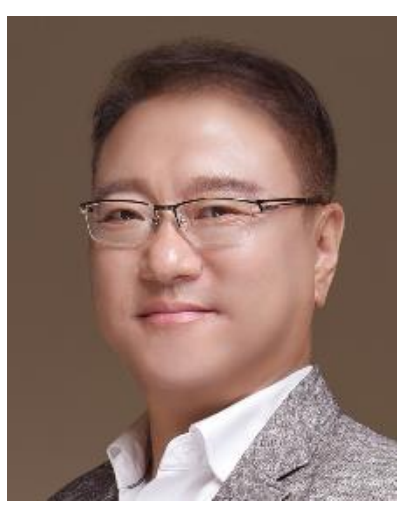

**Yong-Hwa Park** received BS, MS, and PhD in Mechanical Engineering from KAIST in 1991, 1993, and 1999, respectively. In 2000, he joined to Aerospace Department at the University of Colorado at Boulder as a research associate. From 2003-2016, he worked for Samsung Electronics in the Visual Display Division and Samsung Advanced Institute of Technology (SAIT) as a Research Master in the field of micro-optical systems with applications to imaging and display systems. From 2016, he joined KAIST as professor of NOVIC+ (Noise & Vibration Control Plus) at the Department of Mechanical Engineering devoting to research on vibration, acoustics, vision sensors, and condition monitoring with AI.

His research fields include structural vibration; condition monitoring from sound and vibration using AI; health monitoring sensors; and 3D sensors, and lidar for vehicles and robots. He is the conference chair of MOEMS and miniaturized systems in SPIE Photonics West since 2013. He is a vice-president of KSME, KSNVE, KSPE, and member of IEEE and SPIE